\def\grs{GRS~$1915$+$105$}
\def\X1550{XTE~J1550$-$564}

\def\ergcms{erg~cm$^{-2}$~s$^{-1}$}
\def\sx{SIMBOL-X}

\documentclass{mem}
\usepackage{natbib}
\usepackage{txfonts}
\usepackage{balance}
\usepackage{graphicx,epsf,epsfig}
\usepackage[a4paper]{hyperref}
\idline{00}{00}
\begin{document}

\title{A SIMBOL-X View of Microquasars}

\author{J\'er\^ome Rodriguez}

  \offprints{J. Rodriguez}

\institute{Laboratoire AIM, CEA/DSM - CNRS - Universit\'e Paris Diderot, DAPNIA/SAp,
 F-91191 Gif-sur-Yvette, France
\email{jrodriguez@cea.fr}
}

\authorrunning{J. Rodriguez}

\titlerunning{Microquasars seen with SIMBOL-X}

\abstract{Based on spectral simulations, I show how focusing of 
the X-ray radiations above 10 keV will open a new window for the study 
of microquasars. With simulations of soft and hard state spectra of Galactic sources, 
I discuss how \sx\ can help to precisely measure the spin of black holes. 
Spectral study on short $\sim1$~s time scales will also allow the accretion-ejection 
connections to be accessed, and the formation of jets possibly witnessed in   
X-rays. I then turn to external galaxies, and demonstrate that spectral studies of 
hard sources will be possible up to at least $\sim1$ Mpc. For such sources, subtle 
spectral signatures (e.g. a reflection bump) 
will clearly be detected. I finally discuss the implications that these original results 
will bring on the physics of microquasars and black holes.
}
\maketitle{}

\section{Introduction}
Microquasars are X-ray binaries (XRB) in which ejections of material occur. They, 
therefore, are powered by the accretion of matter onto a central compact 
object (either a black hole, BH, or a neutron star). Their behaviour at X-ray energies 
is not different than the behaviour seen in other XRBs. The jets in those systems
can occur under different forms: 
discrete ejections at speeds close to that of the light \citep[e.g.][]{mirabel94}, or  
they can form a more compact jet \citep[e.g.][]{fuchs03}. 
The quite short time scales of the accretion and 
ejection events (as compared to the quasars for example) make them excellent laboratories 
to study the accretion-ejection connections and the physics underlying those phenomena.\\
\indent The X-ray spectra of microquasars/XRBs are usually composed of two 
main components: a thermal one peaking around 1~keV, and a hard X-ray tail 
\citep[e.g.][for a review]{remillard06}. The first is attributed to an accretion 
disc surrounding the compact object, and fed by the companion star. 
The latter emission is usually attributed to the inverse Comptonization of 
the disc photons on a population of hot electrons (thermalized or not), although alternative 
explanations have recently been proposed \citep[e.g.][]{markoff05}. On top of these 
main processes, signatures of fluorescent emission from iron around 6.5 keV or 
refection of hard photons on the accretion disc around 10--50 keV can also be seen.\\
\indent Precise descriptions of the \sx\ mission and the capabilities of the different
 instruments can be found in different papers in these proceedings (Ferrando 2007, 
 Pareschi 2007, Laurent 2007, Lechner 2007). To summarize, \sx\ is a formation flight 
whose payload is made of a single optics module (on the first satellite) focusing the 
X-rays on two detectors placed 20 m away, one above the second, on the second satellite. 
They both have spectro-imaging capabilities.
 It covers the $\sim$0.5 to $\sim$100 keV energy range. In their current versions, the 
Macro-Pixel Detector (MPD) is the upper detection layer and covers the $\sim0.5$--15 keV range, 
while the lower detection layer 
the Cd(Zn)Te detector (CZT) covers the $\sim5$--$100$ keV energy ranges.
Note that all the simulations presented in this paper were done with the set of response 
files\footnote{MPD\_Ref2.rmf and MPD\_TB\_200\_OA0.arf for the soft X-ray detector MPD, 
and CZT\_Ref2.rmf and CZT\_TB\_200\_OA0.arf for the hard X-ray detector CZT.} and scripts provided 
by the organisers of the conference, under {\tt{XSPEC v11.3.2ad}}. All spectra have further
been rebinned in order not to be oversampled and to obtain a good statistics on each spectral 
bin.\\ 
\indent In this paper, I present some results that will be obtained with \sx, concentrating 
more particularly on points that are only achievable with the focusing of the hard
X-rays, and the combination of excellent capabilities from a few tenth of keV 
to $\sim100$~keV: the good angular resolution, extremely low background  and 
high sensitivity of the detectors. In the next section, I study two aspects that will be 
possible with \sx\ concerning two Galactic microquasars, namely \X1550\ and \grs. 
In the third section, I study the possibility of establishing broad band 
X-ray spectra of XRBs from external galaxies.  All these original results and the 
physical implications expected from these studies will be discussed in the last part of 
the paper.\\

\section{Bright sources: microquasars in our own Galaxy}
\subsection{Introduction on spectral states}
Transient microquasars are detected when entering into an outburst. Based 
on the shape of the spectra, the relative contributions of the emitting media, 
the level of rapid ($<1$~s) variability, the presence and type of quasi-periodic 
oscillations, one can distinguish several spectral states 
\citep[e.g.][for precise definitions]{homan05,remillard06}. Two canonical ones are
the Soft State (SS, also referred to as the thermal dominant state) and the hard state (LHS).
In the former the X-ray spectra are dominated by the emission from a thermal disc, and 
the power law tail is soft (its index $\Gamma$ is usually greater than 2) and has almost 
no contribution (less than $25\%$ of the total 2--20 keV unabsorbed flux) to the spectrum
 \citep[for example, in the classification of][]{remillard06}. In the latter the disc 
 is cold ($\sim0.1$ keV), its contribution to the 2--20 keV unabsorbed flux less than 20\%, 
 and $1.4<\Gamma<2.1$. An exponential cut-off is also often observed at energies 20--100 keV.\\
 \indent As these states already correspond to bright phases of an outburst, there is no 
doubt about the feasibility of their study with \sx\ as exemplified by the huge amount of 
archival data from any X-Ray satellite that is available today. The great advantage 
of having a very sensitive telescope above 10 keV brings, however, very interesting
possibilities for the study of BHs in particular, as I show below.

\subsection{States dominated by the thermal component: the thermal state}
 It is thought than during the SS, the disc is 
a "pure" $\alpha$-disc, and that it reaches the last stable orbit around the BH. Hence if 
we are able to measure its value, we will be able to access the spin of the BH which is a key 
parameter to the understanding of those objects. While at first sight current observatories (e.g. 
RXTE, INTEGRAL, Chandra or XMM-Newton) may be able to do it, they either lack sufficient 
sensitivity at soft X-rays (RXTE, INTEGRAL) or lack hard X-ray telescopes (XMM, Chandra).
A recent controversy on the value of the spin of \grs\ \citep{clintock06, middleton06} has shown 
how important is the identification of "pure" $\alpha$-disc states.\\
\indent The advantage of \sx\ over XMM-Newton or Chandra (besides the high degree of pileup in 
those instruments) is the wider energy coverage. Indeed to precisely access the parameters of the disc 
component during the SS, precise estimates of the power law parameters are needed to avoid as much
as possible mixing of the two components. This is done by extending the energy range to a domain
where the disc emission is completely negligible, i.e. above $\sim$20 keV, and by detecting 
the faintest tails.\\
\indent To perform the feasibility of such a study and its pertinence I used the published 
parameters of a SS of the microquasar \X1550 during its 1998 outburst \citep{sobczak99}, as seen 
with RXTE. This observation took place on November 4$^{th}$, 1998. It was choosen because the 
spectrum was so soft (the disc accounted for 92\% of the total 2--20 keV flux) 
that the source was not detected above 20 keV with RXTE/HEXTE \citep{sobczak99}.
\begin{figure}[htbp]
\caption{1~ks simulated MPD+CZT photon spectrum of \X1550 during a SS. The 
best fitted model is superposed as a line, while the individual components 
are represented with a dash-dotted line.} 
\epsfig{file=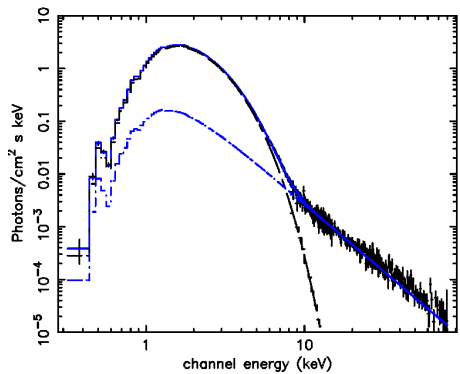,width=\columnwidth}
\label{fig:1550SS}
\end{figure}
Fig. \ref{fig:1550SS} shows the MPD and CZT spectra simulated with an accumulation time of 1~ks. The 
source is clearly detected up to 80~keV. Further simulations showed that it was even detected 
when accumulating the spectra on 100~s only. The best fit model is superposed to the spectra. It consists
of a disc component, a power law and an iron emission line modified by interstellar absorption. All 
these components are required to obtain a good fit. The iron edge component that was input in the simulation
is, however, not required in the fit. All other spectral parameters are compatible with those 
reported in \citet{sobczak99}, and even more precise. In particular the photon index is 
2.41$_{-0.04}^{+0.03}$ (vs. 2.36$\pm0.06$ with RXTE) and the disc radius R$_{in}=46.3\pm0.2$~km
 (vs. 46.4$\pm0.6$ km). The fit on a broader band spectrum helped in increasing the precision
 of the spectral parameters.

\subsection{Rapid X-ray variations: the possibility of spectral studies on a second time scale}
\indent Systematic multi wavelength monitorings of microquasars in outburst have revealed 
a clear but complex association between the accretion seen at X- and Gamma-ray energies, 
and the jets seen in radio/infra red. This has first been seen  in \grs\ 
\cite[e.g.][]{mirabel98}, with ejections of \"bubbles\" following sequences of 
dips terminated by spikes at X-ray energies. This result has 
recently been generalised: in \grs\ ejections of \"bubbles\" will always occur provided 
a sequence of a hard X-ray dip
(i.e. a dip during which the X-ray spectrum is dominated by the power law tail, and the disc 
is cold) of a duration longer than 100~s ended by a short spike takes place \citep{rod06,rod07}. 
Such sequences are dubbed "cycles" due to their repeated occurences. 
\citet{rod06,rod07b} have suggested, based on a spectral analysis of these sequences, that 
the ejected material was the corona. While their analysis makes use of data above 20 keV 
for the first time, they accumulated spectra over times on which the source spectrum 
may evolve, typically tens of second.\\
\begin{figure*}[htbp]
\caption{{\bf{Left:}} 3--13 keV (JEM-X) light
curve of \grs\ dip-spike during a cycle. {\bf{Middle:}} 1~s 
MPD+CZT photon spectra of the peak labeled 1 in the left panel. {\bf{Right:}} Same as in the
middle panel for the dip labeled 2 in the left panel. In both latter cases the best fit model
and individual components are superposed as lines.}
\hspace*{-0.5cm}
\epsfig{file=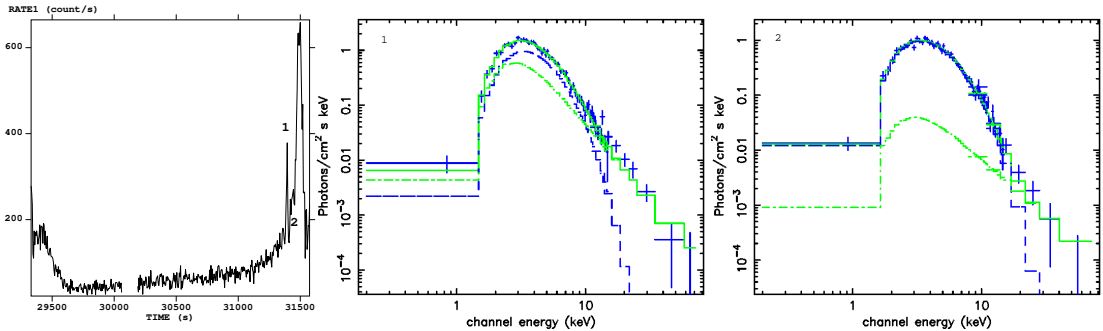,width=15cm}
\label{fig:nu}
\end{figure*}
\indent \sx\ spectra were simulated using the parameters obtained by 
\citet{rod06} from their INTEGRAL observations at two moments during a 
cycle as illustrated in Fig. \ref{fig:nu} left panel. The integration 
time of the simulations were 1~s for each spectra. The spectra were fitted 
with simple a model of a thermal disc and a powerlaw modified by absoprtion. 
The spectral fits to those 
spectra yielded $kT_{\mathrm disc}=1.48\pm0.09$~keV, and $R_{\mathrm in}= 380 \pm60$~km
$\Gamma$=2.8$\pm$0.4 at moment 1,and $kT_{\mathrm disk}=1.74\pm0.07$~keV, 
R$_{\mathrm in}= 255_{-12}^{+40}$~km, and $\Gamma= 2.1\pm1.2$ at moment 2. 
In addition during moment 1 the fluxes of the infdividual components followed 
F$_{\mathrm disc}^{1-100 {\mathrm  keV}}\sim$F$_{\mathrm pl}^{1-100 {\mathrm
keV}}$, while in moment 2, F$_{\mathrm disc}^{1-100 {\mathrm  keV}}\sim10 \times$F$_{\mathrm  pl}^{1-100 {\mathrm keV}}$. 

\section{XRBs in external galaxies}
\indent With the advent of high imaging resolution and high sensitivity X-ray telescopes,
such as XMM-Newton and Chandra, many XRBs have aso been detected in other Galaxies, and 
their Soft X-ray (0.1--10 keV) spectral behaviour studied. The limited energy range has, 
however, hampered a complete study of state changes in those sources, and therefore has 
limited the study of the interplay between the different emitting media in sources found 
in other Galaxies. This topics is more developped in other papers in these
proceedings, especially from the imaging point of view. I briefly show here what 
importance can have a very sensitive hard X-ray detector, and assume that the
simulated source does not suffer from confusion with other sources. Here again 
X-ray telescopes as Chandra and XMM-Newton have shown that many XRBs (and other 
ultra luminous X-ray sources) wer active in other Galaxies. Being sensitive in the
soft X-rays, however, they detected mostly sources that have their bulk of energy 
in that band (i.e. under 10 keV) or that are very bright. In addition to continue to
study those, \sx\  will allow us 
to study sources that have their bulk of emission above 10~keV 
(in the LHS for example), allowing us to precisely follow spectral state changes, 
better characterise source population by adding the less luminous sources to the 
populations, and either discover intrinsically absorbed sources in external 
galaxies.\\
\indent Rather than just a $3-\sigma$ detection (over the whole energy range of \sx)
I tried to see until approximately what distance a decent spectrum could be obtained.
To do so I used spectral parameters similar to those of \X1550\ while in the 
LHS as seen during its 2000 outburst \citep{rod03}, i.e. a cutoff powerlaw 
with $\Gamma=1.49$ and a cut-off  at $33.8$~keV for an intrinsic 0.1--100 keV 
luminosity of 7$\times10^{37}$~\ergcms. The spectra were simulated at two typical 
distances and absorption columns, that of LMC (at 50 kpc, N$_{\mathrm H}=0.682\times10{21}$
cm$^{-2}$) and that of M33 (795 kpc, N$_{\mathrm H}=0.558\times10{21}$
cm$^{-2}$), for respective
integration times of 10~ks for the former and 50~ks for the latter.
\begin{figure}[htbp]
\caption{50 ks joint MPD+CZT spectrum of a LHS source in M33. The lines represent a 
simple absorbed power law model. A significant deviation with the spectrum can 
be seen above $\sim20$~keV.}
\epsfig{file=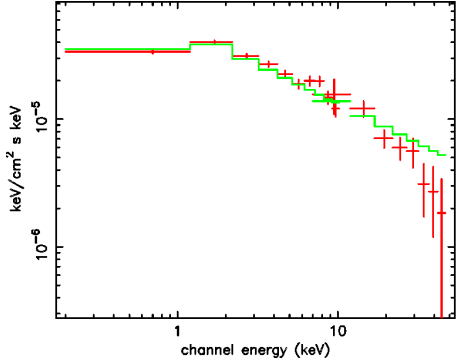,width=\columnwidth}
\label{fig:m33lhs}
\end{figure}
Fig. \ref{fig:m33lhs} shows the result in the case of the M33 source.
While fitting the joint MPD+CZT spectrum with a simple absorbed power law shows
residual above $\sim20$ keV, adding a cut-off improves the fit significantly. An 
F-test indicates a chance probability of $5\times10{-5}$. The spectral parameters are
quite accurate and close to the feed parameter for the simulation:  
N$_{\mathrm H}=0.05±0.02\times10^{22}$ cm$^{-2}$, E$_{\mathrm cut}=33_{-10}^{+24}$~keV, and 
$\Gamma= 1.5\pm0.1$. Beside the detection of hard sources, the detection of a 
spectral break in the energy spectrum of an external XRB will be the first. \\
\begin{figure}[htbp]
\caption{10~ks joint MPD+CZT spectrum of a LHS source in the LMC, with a reflection feature. 
The reflection bump at $\sim20$~keV is obvious.}
\epsfig{file=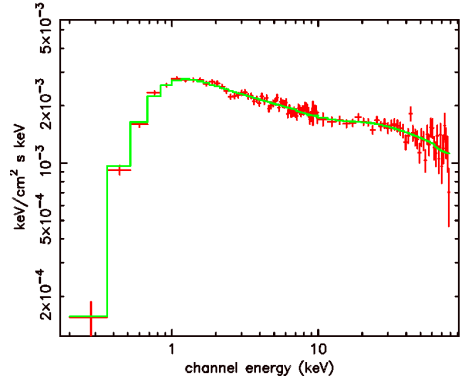,width=\columnwidth}
\label{fig:reflec}
\end{figure}
\indent The last point and the important constraints on  the models of emission led 
me to study whether or not more subtle features visible in the X-ray spectra above 10~keV
will be seen with \sx\ in external galaxies. I simulated a 10~ks exposure Cyg X--1 like spectrum 
for a source in the LMC (thermal Comptonisation with $kT_{\mathrm inj}=0.2$~keV, $kT_{\mathrm e}=67$~keV, and  reflection 
with a scaling factor 0.25, for a 0.1--100 keV flux of $\sim2.3\times10{-10}$~\ergcms). 
The joint MPD+CZT simulated spectrum and the best fit model is reported in Fig. \ref{fig:reflec}.
A simple absorbed Comptonised component represents the spectrum rather well, although
some deviation in the 15--30 keV region can be seen. Adding a reflection component improves
the fits with an F-Test probability of $8\times10^{-9}$ chance improvement. The fitted 
parameters are close to the feed parameters of the simulation with 
$N_{\mathrm H} = 0.05\pm0.02\times 10^{22}$~cm$^{-2}$ a reflection coefficient of $0.22\pm0.06$, 
$kT_{\mathrm inj}=0.21_{-0.02}^{+0.03}$~keV, $kT_{\mathrm e}= 59.51_{-9}^{+13}$~keV. The reflection bump 
is evident in the spectrum (Fig. \ref{fig:reflec}).

\section{Conclusions}
The great advantage of \sx\ compared to the current X-ray satellites is that in addition 
it couples excellent spectro-imaging capabilities below and above 10 keV. As I showed in the 
previous sections, this will allow us to access some physics of accreting compact objects that
is still subject to debate, poorly known or that has even never been studied so far.
\begin{itemize}
\item The detection of very weak corona during the SS in Galactic sources, will enable us to 
obtain the most accurate possible spectral parameters for the accretion disk, and hence access 
one of the key physical parameters for a BH: its spin. The actual controversy, examplified by 
the case of \grs\ \citep{middleton06,clintock06}, shows how crucial the determination of the disc
parameters is.
\item The possibility of performing time resolved spectroscopy on short ($\sim1$~s) time scale
will permit us to precisely follow the evolution of the source immediately after the jet has 
been launched in \grs\ like sources. We may thus witness the formation of the ejecta at X-ray
energies. Such sources present the great advantage to vary very 
rapidely, and show repeated occurences of accretion-ejection linked processes. Hence, the chance 
of observing such phenomena is greater than in any other sources. 
\item In the case of external galaxies, a completely new window will be opened. By detecting hard
sources, non-biased population study will be rendered possible. As showed here it should be
possible to precisely characterise the spectral state up to about 1 Mpc using the $>10$ keV
domain for the first time.
\item The possibility of performing phenomenological as well as physical fits will allow us 
to access the physics of accreting sources in other galaxies, in order to know whether 
they show similar behviour than in our Galaxy (e.g. outbursting sources), and search for  
possible relations with the type of the galaxy. 
\item Studying BH in other galaxies will also increase the number of such sources (only 
20 are confirmed BH in our Galaxy \citet{remillard06}) which will help us constrain the physics
of accretion, and that of BHs.
\end{itemize}
Clearly \sx\ will have a major role in  future studies of accreting-ejecting objects. By 
opening a complete new window in the hard X-rays, it will allow us to answer key questions 
on BHs, but also on the physics of accretion and its link with the ejections. Precise
population studies will be possible while unveiling sources that are faint below 10 keV and emit
their bulk of emission above 10 keV. Additionally the hard X-rays emitted by quiescent Galactic
sources will also be accessed for the first time, with great implication for the emission models,
and sources evolution through outbursts.

\begin{acknowledgements}
I acknowledge S. Corbel and M. Tagger for careful reading and useful 
comments on early versions of this paper. 
\end{acknowledgements}

\bibliographystyle{aa}

\begin{thebibliography}{}

\bibitem[Corbel et al.(2003)]{corbel03} Corbel, S., Nowak, M. A., Fender, R. P., Tzioumis,
 A. K., Markoff, S. 2003, A\&A, 400, 1007

\bibitem[Fender et al.(2004)Fender, Belloni \& Gallo]{fender04} Fender, R.P, Belloni, T., 
\& Gallo, E. 2004, MNRAS, 355, 1105

\bibitem[Ferrando(2007)]{ferrando07}Ferrando, P. 2007, these proceedings

\bibitem[Fuchs et al.(2003)]{fuchs03} Fuchs, Y., Rodriguez, J., Mirabel, F., et al. 2003, 
A\&A, 409, L35.

\bibitem[Gallo(2006)]{gallo06} Gallo, E. 2006,  in ``Proceedings of the VI Microquasar Workshop:
Microquasars and Beyond'',  Eds T. Belloni, PoS(MQW6)009
\bibitem[Homan \& Belloni(2005)]{homan05} Homan, J. \& Belloni, T. 2005, Ap\&SS, 300, 107

\bibitem[Laurent(2007)]{laurent2007}Laurent, P. 2007, these proceedings

\bibitem[Lechner(2007)]{lechner2007}Lechner, P. 2007, these proceedings
\bibitem[McClintock et al.(2006)]{clintock06} McClintock, J.E., Shafee, R, Narayan, R., 
Remillard, R.A., Davis, S.W., Li L. 2006, ApJ, 652, 518.

\bibitem[Markoff, Nowak \& Wilms(2005)Markoff et al.]{markoff05} Markoff, S., Nowak, M. A.
, \& Wilms, J. 2005, ApJ, 635, 1203.
\bibitem[Middleton et al.(2006)]{middleton06}Middleton, M., Done, C., Gierli\'nski, M., 
Davis, S.W. 2006, MNRAS, 373, 1004
\bibitem[Mirabel \& Rodr\'\i guez(1994)]{mirabel94} Mirabel, I.F. \& Rodr\'\i guez, L.F. 1994,
 Nature, 371, 46

\bibitem[Mirabel et al.(1998)]{mirabel98} Mirabel, I.F., Dhawan, V., Chaty, S., et al. 1998, A\&A, 330, L9
\bibitem[Pareschi(2007)]{pareschi}Pareschi, G. 2007, these proceedings

\bibitem[Remillard \& McClintock(2006)]{remillard06} Remillard, R.A. \& McClintock, J.E. 2006, ARA\&A, 44, 49
\bibitem[Rodriguez et al.(2003) Rodriguez, Corbel \& Tomsick]{rod03} Rodriguez, J., Corbel
, S. \& Tomsick, J.A. 2003, ApJ, 595, 1032.
\bibitem[Rodriguez et al.(2006)]{rod06} Rodriguez, J., Pooley, G., Hannikainen, D.C., Lehto, H.J.,
Belloni, T., Cadolle-Bel, M., Corbel, S.   2006, in ``Proceedings of the VI Microquasar Workshop:
Microquasars and Beyond'',  Eds T. Belloni, PoS(MQW6)024
\bibitem[Rodriguez et al.(2007a)]{rod07} Rodriguez, J., Hannikainen, D.C., Shaw, S.E., et al. 2007a 
submitted to ApJ.
\bibitem[Rodriguez et al.(2007b)]{rod07b} Rodriguez, J., Shaw, S.E., Hannikainen, D.C., et al. 2007b 
submitted to ApJ.
\bibitem[Sobczak et al.(1999)]{sobczak99} Sobczak, G.J., McClintock, J.E., Remillard, R.A., et al. 1999, ApJL,
517, 121
\end{thebibliography}

\end{document}